\documentclass[twocolumn,letterpaper,aps,prl,superscriptaddress,showpacs,floatfix]{revtex4}

\usepackage{graphicx}	
\usepackage{xspace}	

\newcommand{\pt}{\mbox{$p_T$}\xspace}
\newcommand{\kt}{\mbox{$k_T$}\xspace}

\newcommand{\Ncoll}{\mbox{$N_{\rm coll}$}\xspace}

\newcommand{\sqsntwo}{\mbox{$\sqrt{s_{_{NN}}}=200$~GeV}\xspace}

\newcommand{\rda}{\mbox{$R_{dA}$}\xspace}
\newcommand{\pp}{\mbox{$p$$+$$p$}\xspace}
\newcommand{\dau}{\mbox{$d$$+$Au}\xspace}

\newcommand{\auau}{\mbox{Au$+$Au}\xspace}
\newcommand{\cucu}{\mbox{Cu$+$Cu}\xspace}
\newcommand{\jpsi}{\mbox{$J/\psi$}\xspace}

\begin{document}

\title{Cold-nuclear-matter effects on heavy-quark production 
at forward and backward rapidity in $d$$+$Au collisions
at $\sqrt{s_{_{NN}}}=200$~GeV }

\newcommand{\abilene}{Abilene Christian University, Abilene, Texas 79699, USA}
\newcommand{\augie}{Department of Physics, Augustana College, Sioux Falls, South Dakota 57197, USA}
\newcommand{\banaras}{Department of Physics, Banaras Hindu University, Varanasi 221005, India}
\newcommand{\barc}{Bhabha Atomic Research Centre, Bombay 400 085, India}
\newcommand{\baruch}{Baruch College, City University of New York, New York, New York, 10010 USA}
\newcommand{\bnlcoll}{Collider-Accelerator Department, Brookhaven National Laboratory, Upton, New York 11973-5000, USA}
\newcommand{\bnlphys}{Physics Department, Brookhaven National Laboratory, Upton, New York 11973-5000, USA}
\newcommand{\caucr}{University of California - Riverside, Riverside, California 92521, USA}
\newcommand{\charlesczech}{Charles University, Ovocn\'{y} trh 5, Praha 1, 116 36, Prague, Czech Republic}
\newcommand{\chonbuk}{Chonbuk National University, Jeonju, 561-756, Korea}
\newcommand{\ciae}{Science and Technology on Nuclear Data Laboratory, China Institute of Atomic Energy, Beijing 102413, P.~R.~China}
\newcommand{\cns}{Center for Nuclear Study, Graduate School of Science, University of Tokyo, 7-3-1 Hongo, Bunkyo, Tokyo 113-0033, Japan}
\newcommand{\colorado}{University of Colorado, Boulder, Colorado 80309, USA}
\newcommand{\columbia}{Columbia University, New York, New York 10027 and Nevis Laboratories, Irvington, New York 10533, USA}
\newcommand{\czechtech}{Czech Technical University, Zikova 4, 166 36 Prague 6, Czech Republic}
\newcommand{\dapnia}{Dapnia, CEA Saclay, F-91191, Gif-sur-Yvette, France}
\newcommand{\elte}{ELTE, E{\"o}tv{\"o}s Lor{\'a}nd University, H - 1117 Budapest, P{\'a}zm{\'a}ny P. s. 1/A, Hungary}
\newcommand{\ewha}{Ewha Womans University, Seoul 120-750, Korea}
\newcommand{\fit}{Florida Institute of Technology, Melbourne, Florida 32901, USA}
\newcommand{\fsu}{Florida State University, Tallahassee, Florida 32306, USA}
\newcommand{\gsu}{Georgia State University, Atlanta, Georgia 30303, USA}
\newcommand{\hanyang}{Hanyang University, Seoul 133-792, Korea}
\newcommand{\hiroshima}{Hiroshima University, Kagamiyama, Higashi-Hiroshima 739-8526, Japan}
\newcommand{\ihepprot}{IHEP Protvino, State Research Center of Russian Federation, Institute for High Energy Physics, Protvino, 142281, Russia}
\newcommand{\illuiuc}{University of Illinois at Urbana-Champaign, Urbana, Illinois 61801, USA}
\newcommand{\inrras}{Institute for Nuclear Research of the Russian Academy of Sciences, prospekt 60-letiya Oktyabrya 7a, Moscow 117312, Russia}
\newcommand{\instpasczech}{Institute of Physics, Academy of Sciences of the Czech Republic, Na Slovance 2, 182 21 Prague 8, Czech Republic}
\newcommand{\isu}{Iowa State University, Ames, Iowa 50011, USA}
\newcommand{\jaea}{Advanced Science Research Center, Japan Atomic Energy Agency, 2-4 Shirakata Shirane, Tokai-mura, Naka-gun, Ibaraki-ken 319-1195, Japan}
\newcommand{\jyvaskyla}{Helsinki Institute of Physics and University of Jyv{\"a}skyl{\"a}, P.O.Box 35, FI-40014 Jyv{\"a}skyl{\"a}, Finland}
\newcommand{\kek}{KEK, High Energy Accelerator Research Organization, Tsukuba, Ibaraki 305-0801, Japan}
\newcommand{\korea}{Korea University, Seoul, 136-701, Korea}
\newcommand{\kurchatov}{Russian Research Center ``Kurchatov Institute", Moscow, 123098 Russia}
\newcommand{\kyoto}{Kyoto University, Kyoto 606-8502, Japan}
\newcommand{\labllr}{Laboratoire Leprince-Ringuet, Ecole Polytechnique, CNRS-IN2P3, Route de Saclay, F-91128, Palaiseau, France}
\newcommand{\lahorelums}{Physics Department, Lahore University of Management Sciences, Lahore, Pakistan}
\newcommand{\lawllnl}{Lawrence Livermore National Laboratory, Livermore, California 94550, USA}
\newcommand{\losalamos}{Los Alamos National Laboratory, Los Alamos, New Mexico 87545, USA}
\newcommand{\lpc}{LPC, Universit{\'e} Blaise Pascal, CNRS-IN2P3, Clermont-Fd, 63177 Aubiere Cedex, France}
\newcommand{\lund}{Department of Physics, Lund University, Box 118, SE-221 00 Lund, Sweden}
\newcommand{\maryland}{University of Maryland, College Park, Maryland 20742, USA}
\newcommand{\mass}{Department of Physics, University of Massachusetts, Amherst, Massachusetts 01003-9337, USA }
\newcommand{\michigan}{Department of Physics, University of Michigan, Ann Arbor, Michigan 48109-1040, USA}
\newcommand{\muenster}{Institut fur Kernphysik, University of Muenster, D-48149 Muenster, Germany}
\newcommand{\muhlenberg}{Muhlenberg College, Allentown, Pennsylvania 18104-5586, USA}
\newcommand{\myongji}{Myongji University, Yongin, Kyonggido 449-728, Korea}
\newcommand{\nagasaki}{Nagasaki Institute of Applied Science, Nagasaki-shi, Nagasaki 851-0193, Japan}
\newcommand{\newmex}{University of New Mexico, Albuquerque, New Mexico 87131, USA }
\newcommand{\nmsu}{New Mexico State University, Las Cruces, New Mexico 88003, USA}
\newcommand{\ohio}{Department of Physics and Astronomy, Ohio University, Athens, Ohio 45701, USA}
\newcommand{\ornl}{Oak Ridge National Laboratory, Oak Ridge, Tennessee 37831, USA}
\newcommand{\orsay}{IPN-Orsay, Universite Paris Sud, CNRS-IN2P3, BP1, F-91406, Orsay, France}
\newcommand{\peking}{Peking University, Beijing 100871, P.~R.~China}
\newcommand{\pnpi}{PNPI, Petersburg Nuclear Physics Institute, Gatchina, Leningrad region, 188300, Russia}
\newcommand{\riken}{RIKEN Nishina Center for Accelerator-Based Science, Wako, Saitama 351-0198, Japan}
\newcommand{\rikjrbrc}{RIKEN BNL Research Center, Brookhaven National Laboratory, Upton, New York 11973-5000, USA}
\newcommand{\rikkyo}{Physics Department, Rikkyo University, 3-34-1 Nishi-Ikebukuro, Toshima, Tokyo 171-8501, Japan}
\newcommand{\saopaulo}{Universidade de S{\~a}o Paulo, Instituto de F\'{\i}sica, Caixa Postal 66318, S{\~a}o Paulo CEP05315-970, Brazil}
\newcommand{\seoulnat}{Seoul National University, Seoul, Korea}
\newcommand{\stonybrkc}{Chemistry Department, Stony Brook University, SUNY, Stony Brook, New York 11794-3400, USA}
\newcommand{\stonycrkp}{Department of Physics and Astronomy, Stony Brook University, SUNY, Stony Brook, New York 11794-3400, USA}
\newcommand{\tenn}{University of Tennessee, Knoxville, Tennessee 37996, USA}
\newcommand{\titech}{Department of Physics, Tokyo Institute of Technology, Oh-okayama, Meguro, Tokyo 152-8551, Japan}
\newcommand{\tsukuba}{Institute of Physics, University of Tsukuba, Tsukuba, Ibaraki 305, Japan}
\newcommand{\vandy}{Vanderbilt University, Nashville, Tennessee 37235, USA}
\newcommand{\waseda}{Waseda University, Advanced Research Institute for Science and Engineering, 17 Kikui-cho, Shinjuku-ku, Tokyo 162-0044, Japan}
\newcommand{\weizmann}{Weizmann Institute, Rehovot 76100, Israel}
\newcommand{\wigner}{Institute for Particle and Nuclear Physics, Wigner Research Centre for Physics, Hungarian Academy of Sciences (Wigner RCP, RMKI) H-1525 Budapest 114, POBox 49, Budapest, Hungary}
\newcommand{\yonsei}{Yonsei University, IPAP, Seoul 120-749, Korea}
\affiliation{\abilene}
\affiliation{\augie}
\affiliation{\banaras}
\affiliation{\barc}
\affiliation{\baruch}
\affiliation{\bnlcoll}
\affiliation{\bnlphys}
\affiliation{\caucr}
\affiliation{\charlesczech}
\affiliation{\chonbuk}
\affiliation{\ciae}
\affiliation{\cns}
\affiliation{\colorado}
\affiliation{\columbia}
\affiliation{\czechtech}
\affiliation{\dapnia}
\affiliation{\elte}
\affiliation{\ewha}
\affiliation{\fit}
\affiliation{\fsu}
\affiliation{\gsu}
\affiliation{\hanyang}
\affiliation{\hiroshima}
\affiliation{\ihepprot}
\affiliation{\illuiuc}
\affiliation{\inrras}
\affiliation{\instpasczech}
\affiliation{\isu}
\affiliation{\jaea}
\affiliation{\jyvaskyla}
\affiliation{\kek}
\affiliation{\korea}
\affiliation{\kurchatov}
\affiliation{\kyoto}
\affiliation{\labllr}
\affiliation{\lahorelums}
\affiliation{\lawllnl}
\affiliation{\losalamos}
\affiliation{\lpc}
\affiliation{\lund}
\affiliation{\maryland}
\affiliation{\mass}
\affiliation{\michigan}
\affiliation{\muenster}
\affiliation{\muhlenberg}
\affiliation{\myongji}
\affiliation{\nagasaki}
\affiliation{\newmex}
\affiliation{\nmsu}
\affiliation{\ohio}
\affiliation{\ornl}
\affiliation{\orsay}
\affiliation{\peking}
\affiliation{\pnpi}
\affiliation{\riken}
\affiliation{\rikjrbrc}
\affiliation{\rikkyo}
\affiliation{\saopaulo}
\affiliation{\seoulnat}
\affiliation{\stonybrkc}
\affiliation{\stonycrkp}
\affiliation{\tenn}
\affiliation{\titech}
\affiliation{\tsukuba}
\affiliation{\vandy}
\affiliation{\waseda}
\affiliation{\weizmann}
\affiliation{\wigner}
\affiliation{\yonsei}
\author{A.~Adare} \affiliation{\colorado}
\author{C.~Aidala} \affiliation{\losalamos} \affiliation{\mass} \affiliation{\michigan}
\author{N.N.~Ajitanand} \affiliation{\stonybrkc}
\author{Y.~Akiba} \affiliation{\riken} \affiliation{\rikjrbrc}
\author{R.~Akimoto} \affiliation{\cns}
\author{H.~Al-Bataineh} \affiliation{\nmsu}
\author{H.~Al-Ta'ani} \affiliation{\nmsu}
\author{J.~Alexander} \affiliation{\stonybrkc}
\author{K.R.~Andrews} \affiliation{\abilene}
\author{A.~Angerami} \affiliation{\columbia}
% ...
\collaboration{PHENIX Collaboration} \noaffiliation

\date{\today}


\begin{abstract}

The PHENIX experiment has measured open heavy-flavor production 
via semileptonic decay 
over the transverse momentum range $1<p_T<6$~GeV/$c$ at forward and backward rapidity 
($1.4<|y|<2.0$) in $d$$+$Au and $p$$+$$p$ collisions at 
$\sqrt{s_{_{NN}}}=200$~GeV. 
In central $d$$+$Au collisions an enhancement of 
heavy-flavor muon production is observed at backward rapidity, whereas 
suppression is seen at forward rapidity relative to the yield in 
$p$$+$$p$ collisions scaled by the number of binary collisions.
The difference observed between forward and backward rapidity 
exceeds predictions based on a model of initial parton density modification. 
These results can be used to probe predicted cold nuclear matter 
effects, which may significantly affect heavy-quark production at the 
Relativistic Heavy Ion Collider and the Large Hadron Collider, in addition to 
helping constrain the magnitude of charmonia breakup effects in 
nuclear matter.

\end{abstract}

	

\maketitle


Heavy quarks are essential probes of the evolution of the medium created 
in heavy-ion collisions, because they are produced in the early stages 
of nuclear collisions. Heavy-quark production has been measured via 
semileptonic decay electrons and muons, as well as fully reconstructed 
$D$ mesons, at RHIC and the LHC~\cite{aliceD:2012,aliceD:2013}. In \pp collisions, heavy-quark 
production tests perturbative quantum chromodynamics and provides 
a baseline for heavy-ion collisions~\cite{ppg057,ppg065,starpp:2011}. In 
central \auau collisions at \sqsntwo, strong suppression of high 
transverse momentum (\pt) electrons from semileptonic decay 
of open heavy-flavor hadrons has been observed at 
midrapidity~\cite{ppg056,ppg066}. At forward 
rapidity, a similar level of suppression has been measured for the 
production of heavy-flavor muons in central \cucu 
collisions~\cite{ppg117}. Although suppression of high \pt particles was 
predicted as an effect of partonic energy loss in the dense medium 
created in heavy-ion 
collisions~\cite{mustafa:2005,moore:2005,hees:2006}, it is difficult to 
account for this comparable suppression solely with hot nuclear matter 
effects~\cite{ppg117,Djordjevic:2006}.  To interpret such measurements, it is essential 
to probe underlying cold-nuclear-matter (CNM) effects, which may also be 
present.

Control experiments with \dau collisions allow us to probe those CNM 
effects, including modifications of the parton distribution function (PDF) and 
\kt broadening, with minimal impact from the hot nuclear 
medium. Because heavy quarks are produced primarily by gluon fusion at 
RHIC, modification of the gluon density in the nucleus can be observed in the charm 
and bottom production rates~\cite{eps09:2009,eps09s:2012}. Based on 
{\sc pythia}~\cite{pythia:2006} calculations, the average parton momentum 
fraction $x$ in the Au nucleus leading to heavy-flavor muons with 
$1<p_{T}^{\mu}<6$ GeV/$c$ at backward ($-2.0<y<-1.4$, Au-going 
direction) and forward ($1.4< y<2.0$, $d$-going direction) rapidity is 
$\approx8\times10^{-2}$ and $\approx5\times10^{-3}$ for 
the antishadowing and shadowing regions, respectively. Parton energy loss and 
multiple scattering in the nucleus can change the resulting heavy-flavor 
hadron momentum spectrum~\cite{vitev:2007}. Previous results in \dau 
collisions at midrapidity show a significant enhancement of heavy-flavor 
electrons at moderate \pt~\cite{ppg131}. In this Letter, we present 
measurements of the \pt spectra and the nuclear modification factor (\rda) 
of negatively charged muons from open heavy flavor at forward and backward 
rapidity in \dau collisions at \sqsntwo.

The \dau and \pp data presented here were recorded with the PHENIX 
detector during the 2008 and 2009 RHIC running periods, respectively. 
The minimum-bias collision is selected by using the beam-beam counter 
(BBC)~\cite{bbc:2003}, and this selection covers $88\pm4\%$ ($55\pm5\%$) 
of the total \dau (\pp) inelastic 
cross section~\cite{white:2005}. The integrated luminosity, sampled using 
single muon triggers~\cite{ppg117} in coincidence with the minimum-bias trigger, used for
this analysis of \dau (\pp) collisions is $50~{\rm nb}^{-1}$ ($10~{\rm pb}^{-1}$).    
The \dau collisions are categorized 
into five centrality classes: 0\%--20\%, 20\%--40\%, 40\%--60\%, 
60\%--88\%, and 0\%-100\%, where 0\%--20\% represents the 20\% highest 
multiplicity events, as determined by the amount of total charge 
deposited in the BBC on the Au-going side. For each centrality class, 
the average number of binary nucleon-nucleon collisions $\langle\Ncoll\rangle$ is calculated 
from the BBC charge in a Glauber model~\cite{glauber}. 
Correction for the underlying event correlation 
and the efficiency of the BBC trigger to 100\% is applied as 
in~\cite{ppg125,ppg146}. The values of $\langle\Ncoll\rangle$ for the 
five \dau centrality classes specified above are $15.1\pm1.0$, 
$10.2\pm0.7$, $6.6\pm0.4$, $3.2\pm0.2$, and $7.6\pm0.4$ respectively.

Two muon spectrometers~\cite{phenix:2003} provide full azimuthal 
coverage in the pseudorapidity range $-2.2<\eta<-1.2$ (backward 
rapidity) and $1.2<\eta<2.4$ (forward rapidity). Each muon arm, located 
behind copper (19 cm) and iron (60 cm) absorbers, is composed of a 
muon tracker (MuTr) followed by a muon identifier (MuID). The MuTr 
comprises three stations of cathode strip chambers surrounded by a 
radial magnetic field, and the MuID comprises five interleaved layers 
of steel absorber and Iarocci tube planes. The MuTr provides the 
momentum measurement for charged tracks in the magnetic field. 
The momentum information for each charged track is then combined with 
its penetration depth reported by the MuID to provide effective 
discrimination between muons and hadrons (pion rejection rate: 
$\sim10^{-3}$)~\cite{muonarm:2003}.

Despite the large hadron rejection power of the muon arms and strict 
selection criteria, most of the tracks reaching the last MuID layer are 
not heavy-flavor muons. For $\pt<3~{\rm GeV}/c$ the majority of these 
background tracks originate from the decays of light-flavor mesons 
(mostly $\pi^{\pm}$ and $K^{\pm}$) into muons before reaching the 
absorber material. Another source of background, called ``punch-through 
hadrons," are the hadrons produced at the collision vertex, which  
penetrate all MuID layers. These become the dominant background at 
$\pt>3~{\rm GeV}/c$. Other, less significant sources of background 
include muons from hadrons that decay inside the MuTr which are 
misreconstructed with erroneously high \pt, muons from heavy-flavor 
resonances ($\chi_{c}$, $J/\psi$, $\psi'$, and $\Upsilon$), and muons 
from light vector mesons ($\rho$, $\phi$, and $\omega$). 
The backgrounds are subtracted as follows.

For each data set, we measure the double differential heavy-flavor muon 
invariant yield, defined as
\begin{equation}\label{eq:smyield}
\frac{d^{2}N^{\mu}}{2 \pi \pt d\pt dy} 
= \frac{1}{2\pi \pt\Delta \pt \Delta y}\frac{N_{I}-N_{C}-N_{F}-N_{\jpsi}}
{(N_{\rm evt}/c)A\epsilon}
\end{equation}
where $\Delta p_{T}$ and $\Delta y$ are the bin widths in $\pt$ and $y$; 
$N_{I}$ is the number of inclusive muon candidates; $N_{C}$ is the 
number of decay and punch-through hadron background tracks determined 
using a hadron cocktail method (described below); 
$N_{F}$ is the estimated number of fake tracks that pass the 
selection criteria; $N_{J/\psi}$ is the number of muons from \jpsi 
decays; $N_{\rm evt}$ is the number of sampled events; $A\epsilon$ is 
the detector acceptance and efficiency correction; and $c$ is the BBC 
bias correction factor for the trigger efficiency and centrality determination 
of events containing a heavy-flavor muon. The 
contribution from remaining background sources is less than 
5\%~\cite{ppg077,ppg117}. Only negative 
muons are used, because the signal-to-background 
ratio is better than for positive muons~\cite{ppg117}. 
The typical signal-to-background ratio, 
$N^{\mu}/ (N_{C}+N_{F}+N_{\jpsi})$, increases from 
$0.3$ at $\pt=1~{\rm GeV}/c$ to $0.6$ at $\pt=6~{\rm GeV}/c$. The hadron 
cocktail method estimates the overall background owing to light hadron sources 
using a fully data-driven {\sc geant} simulation based on measured \pt spectra. 
Details on background estimation procedure and associated systematic 
uncertainty are described in~\cite{ppg057,ppg077,ppg117}.

\begin{figure}[t]
\includegraphics[width=1.0\linewidth]{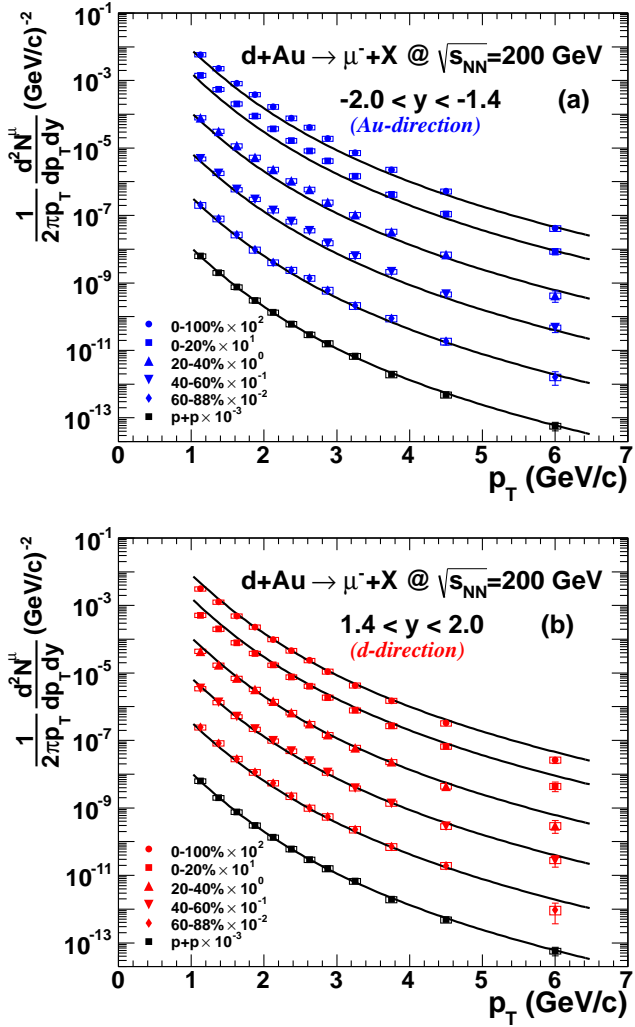}
\caption{(color online). Invariant yield of negatively charged 
heavy-flavor muons as a function of \pt in \pp collisions at \sqsntwo 
(black squares) and in \dau collisions for different centralities at (a) 
backward rapidity (Au-going) and (b) forward rapidity ($d$-going). The 
solid lines represent a fit to the \pp invariant yield, scaled by the 
number of binary collisions, $\langle\Ncoll\rangle$, for each centrality 
class.}
\label{fig:pTspectra}
\end{figure}

Figure~\ref{fig:pTspectra} shows the invariant yield of heavy-flavor 
muons in \dau collisions at backward and forward rapidity along with the 
invariant yield in \pp collisions. The vertical bars represent 
statistical uncertainties, while boxes are systematic uncertainties 
in the acceptance and efficiency correction, 
background estimate, and trigger bias correction for each centrality 
class. The main source of the systematic uncertainties is the background 
estimate including initial hadron production ($\sim$10\%) and hadron 
simulation ($\sim$10\%). All components of the systematic uncertainty 
are added in quadrature. Solid lines show a modified Kaplan function 
$A\big[1+(\pt/8.3~({\rm GeV}/c))^{2}\big]^{-3.9}$~\cite{yoh:1978}, fit to the \pt 
spectrum in \pp collisions, and then scaled by $\langle\Ncoll\rangle$ 
for each \dau centrality class. The \pp results are consistent with 
previous PHENIX measurements~\cite{ppg117}.

\begin{figure}[t]
\includegraphics[width=1.0\linewidth]{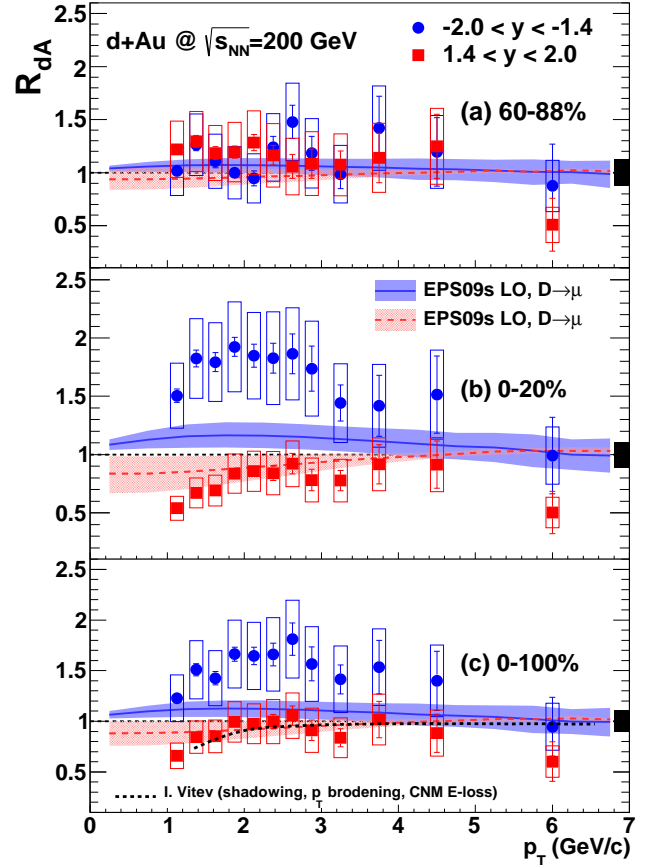}
\caption{(color online). The nuclear modification factor \rda, for 
negatively charged heavy-flavor muons in \dau collisions for the (a) 
60\%--88\%, (b) 0\%--20\%, and (c) 0\%--100\% most central collisions. The black 
boxes on the right side indicate the global scaling uncertainty. The red 
dashed (blue solid) lines in each panel are calculations at forward 
(backward) rapidity based on the EPS09s nPDF set~\cite{eps09s:2012}. The 
theoretical calculation shown in (c) is for forward 
rapidity~\cite{vitev:2007}.}
\label{fig:RdApT}
\end{figure}

To quantify nuclear effects in \dau collisions, we calculate the ratio 
of heavy-flavor muon yields in \dau to \pp collisions scaled by the 
average number of binary collisions for a given centrality bin,
\begin{equation}
R_{dA} = \frac{dN^{\mu^{-}}_{dA}/dp_T}
{\langle N_{\rm coll} \rangle \times dN^{\mu^{-}}_{pp}/dp_T}.
\label{eqn:rda}
\end{equation}
Figure~\ref{fig:RdApT} shows \rda as a function of \pt for heavy-flavor 
muons in different \dau centrality classes. Vertical bars represent the 
statistical uncertainties for $\rda,$ which are the quadratic sum of the 
statistical uncertainties for the invariant yields of \pp and \dau 
collisions.  Boxes around the data points are the systematic 
uncertainties. The global scaling uncertainty  
in $\langle\Ncoll\rangle$ and the BBC efficiency is shown 
as a box centered around unity at the right edge of the plot.

For the most peripheral collisions in both rapidity ranges, \rda shows no 
overall modification. For the most central collisions, a clear 
enhancement is observed at backward rapidity. This enhancement shows 
a \pt dependence consistent with \pt broadening and gluon antishadowing. 
A suppression is observed at forward rapidity in the most 
central collisions. At forward rapidity, \pt broadening is indicated by
the slope of \rda, combined with a suppression that could be 
caused by gluon shadowing and/or partonic energy loss in CNM.

The dotted line in Fig.~\ref{fig:RdApT}(c) is a prediction of \rda 
for muons from $D$ and $B$ mesons at forward rapidity, $y=1.7$~\cite{sharma:2009,vitev:2007}. This prediction, including CNM 
effects such as shadowing, initial-state energy loss, and \kt 
broadening, is consistent with the data at forward rapidity for the 
0\%--100\% centrality class. The same model, with additional energy loss 
in deconfined hot nuclear matter, also describes the forward 
heavy-flavor muon results in central \cucu collisions within 
uncertainties~\cite{ppg117}. This agreement and the suppression at 
forward rapidity in central \dau collisions suggest that CNM effects may 
be important for the interpretation of the suppression of heavy-flavor 
muon production at forward rapidity at RHIC~\cite{ppg117} and the Large 
Hadron Collider~\cite{alicemuon:2012}.

\begin{figure}[t]
\includegraphics[width=1.0\linewidth]{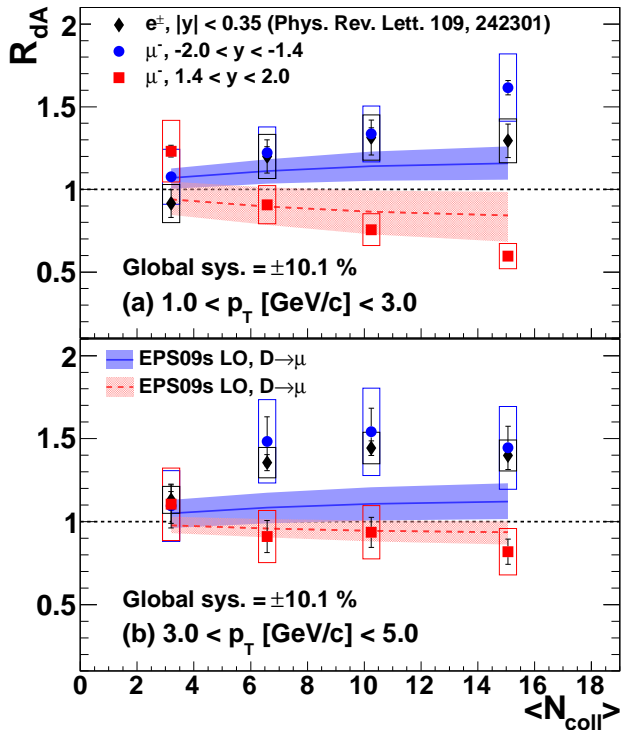}
\caption{(color online). Comparison of \rda as a function of 
$\langle\Ncoll\rangle$ for heavy-flavor leptons 
from different rapidity and \pt bins. Data in the top (bottom) panel are 
from low ($1<\pt~[{\rm GeV}/c]<3$) and moderate ($3<\pt~[{\rm 
GeV}/c]<5$) \pt ranges. Diamonds represent heavy-flavor electrons at 
midrapidity~\cite{ppg131} and squares (circles) represent heavy-flavor 
muons at forward (backward) rapidity.
}
\label{fig:RdANcoll}
\end{figure}

\begin{figure}[t]
\includegraphics[width=1.0\linewidth]{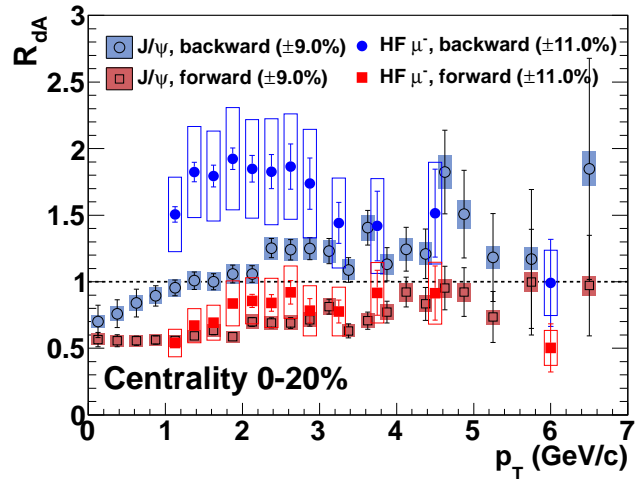}
\caption{(color online). The nuclear modification factor \rda for 
$J/\psi$~\cite{ppg125} and heavy-flavor muons for the 0\%--20\% 
centrality class. The global systematic uncertainty on each distribution is 
shown as a percentage in the legend.
}
\label{fig:RdAJpsi}
\end{figure}

We use the EPS09s leading-order (LO) nuclear PDF (nPDF) 
set~\cite{eps09s:2012} to calculate \rda for muons from $D$ mesons at 
backward (solid lines) and forward (dashed lines) rapidity as 
described in~\cite{nagle:2011}. The EPS09s nPDF further incorporates 
a spatial dependence within the nucleus to the nPDF. The modification of nPDF 
is determined based on the input parameters $x$, momentum transfer 
($Q^{2}$) of charm production generated by 
{\sc pythia}~\cite{pythia:2006}, and transverse radial positions of binary 
collisions in the nucleus for each centrality class. The uncertainty 
bands are calculated as described in~\cite{eps09:2009}. From this calculation, 
we can take solely the initial parton density modification into account. In central 
collisions, shown in Fig.~\ref{fig:RdApT}(b), the EPS09s nPDF based calculation 
does not reproduce the data at backward rapidity, particularly in the 
moderate \pt region; the difference is $\sim2\sigma$ near 
$\pt=2~{\rm GeV}/c$. At forward rapidity, \rda calculated with the EPS09s 
nPDF is consistent with the data over the entire \pt range within 
the systematic uncertainties of the data and calculation. The presence of other 
CNM effects is suggested, because the difference between 
forward and backward rapidity is significantly larger in the data than 
in the EPS09 nPDF calculation.

Figure~\ref{fig:RdANcoll} shows the heavy-flavor muon \rda as a function 
of $\langle\Ncoll\rangle$ for (a) $1.0<\pt~[{\rm GeV}/c]<3.0$ and (b) 
$3.0<\pt~[{\rm GeV}/c]<5.0$, compared to the heavy-flavor electron 
measurement at midrapidity~\cite{ppg131}. Bars (boxes) around the data 
points represent the statistical (systematic) uncertainties determined 
as the quadratic sum of statistical (systematic) uncertainties on \rda 
for each centrality class. In both \pt ranges midrapidity and backward 
rapidity results agree within systematic uncertainties, showing a large 
enhancement for more central collisions. At forward rapidity the low-\pt 
bin shows suppression increasing with centrality, whereas the high-\pt 
bin shows little or no centrality dependence. The EPS09s nPDF based 
calculations are consistent with the data at forward rapidity within 
uncertainties.

Quarkonia and open heavy-flavor hadrons are sensitive to the same 
effects on heavy-quark production. However, quarkonium states are 
additionally influenced by breakup in nuclear matter. Therefore, 
open heavy-flavor production can provide a baseline for interpreting the 
nuclear breakup of quarkonia. Previous measurements suggest that nuclear breakup 
has a significant effect on quarkonia production in nuclear 
collisions~\cite{ppg078,ppg125,ppg151,ppg109,nagle:2011,kopeliovich:2011,lourenco:2009}. 

Figure~\ref{fig:RdAJpsi} shows a comparison of \rda between heavy-flavor 
muons and \jpsi~\cite{ppg125} for central collisions. A similar behavior 
across the entire \pt range is observed at forward rapidity, within the 
systematic uncertainties, whereas a distinct difference is seen at 
backward rapidity, particularly for $\pt<2.5~{\rm GeV}/c$ where charm 
contributions dominate over those from bottom~\cite{ppg094}. The larger 
difference of the \rda between \jpsi and open charm at backward rapidity 
compared to forward rapidity could be related to the longer time this 
$c\bar{c}$ state requires to traverse the nuclear matter or the larger 
density of comoving particles after the initial collision at backward 
rapidity~\cite{vogt:2000}. This comparison suggests that an additional CNM 
effect, nuclear breakup, significantly affects \jpsi production at mid- 
and backward rapidity. This measurement provides a key additional 
constraint on theoretical models attempting to describe quarkonia yields 
in nuclear collisions.

We have presented a measurement of negatively charged 
heavy-flavor muons produced at forward and backward rapidity in \dau 
collisions at \sqsntwo, for several centrality classes. We observe no 
significant modification in the most peripheral \dau collisions. 
However, in central \dau collisions, suppression (enhancement) of 
heavy-flavor muons is observed at forward (backward) rapidity. 
The large difference between forward and backward 
rapidity, which is not reproduced by {\sc pythia} calculations with the EPS09s nPDF 
sets, suggests that various CNM effects combine to produce the observed 
modifications. A comparison between the measured nuclear modification factors for 
\jpsi and open heavy-flavor production provides 
strong indication that nuclear breakup significantly affects quarkonia 
production.




We thank the staff of the Collider-Accelerator and Physics
Departments at Brookhaven National Laboratory and the staff of
the other PHENIX participating institutions for their vital
contributions.  We acknowledge support from the 
Office of Nuclear Physics in the
Office of Science of the Department of Energy, 
the National Science Foundation, 
Abilene Christian University Research Council, 
Research Foundation of SUNY, and 
Dean of the College of Arts and Sciences, Vanderbilt University (U.S.A),
Ministry of Education, Culture, Sports, Science, and Technology
and the Japan Society for the Promotion of Science (Japan),
Conselho Nacional de Desenvolvimento Cient\'{\i}fico e
Tecnol{\'o}gico and Funda\c c{\~a}o de Amparo {\`a} Pesquisa do
Estado de S{\~a}o Paulo (Brazil),
Natural Science Foundation of China (P.~R.~China),
Ministry of Education, Youth and Sports (Czech Republic),
Centre National de la Recherche Scientifique, Commissariat
{\`a} l'{\'E}nergie Atomique, and Institut National de Physique
Nucl{\'e}aire et de Physique des Particules (France),
Bundesministerium f\"ur Bildung und Forschung, Deutscher
Akademischer Austausch Dienst, and Alexander von Humboldt Stiftung (Germany),
Hungarian National Science Fund, OTKA (Hungary), 
Department of Atomic Energy and Department of Science and Technology (India), 
Israel Science Foundation (Israel), 
National Research Foundation and WCU program of the 
Ministry Education Science and Technology (Korea),
Physics Department, Lahore University of Management Sciences (Pakistan),
Ministry of Education and Science, Russian Academy of Sciences,
Federal Agency of Atomic Energy (Russia),
VR and Wallenberg Foundation (Sweden), 
the U.S. Civilian Research and Development Foundation for the
Independent States of the Former Soviet Union, 
the Hungarian American Enterprise Scholarship Fund,
the US-Hungarian Fulbright Foundation for Educational Exchange,
and the US-Israel Binational Science Foundation.


\bibliographystyle{apsrev}


\end{document}